\documentclass[12pt,a4paper]{article}

\usepackage[a4paper,rmargin=1.7cm,lmargin=1.7cm,tmargin=1.0cm,bmargin=1.0cm,includefoot]{geometry}

\usepackage{array}
\usepackage{epsfig}

\usepackage{float}

\usepackage[mathscr]{eucal}

\usepackage{cite}

\usepackage[namelimits]{amsmath}
\usepackage{amssymb}
\usepackage{amsfonts}
\usepackage{bbm}

\usepackage{multirow}
\usepackage{color}

\usepackage[T1]{fontenc}     
\usepackage{ae}              
\usepackage{indentfirst}

\usepackage[affil-it]{authblk}

\begin{document}

\title{\bf Analytic components for the hadronic total cross-section:
Fractional calculus and Mellin transform}

\author[1]{E. Capelas de Oliveira\footnote{capelas@ime.unicamp.br}}
\author[2]{M.J. Menon\footnote{menon@ifi.unicamp.br}}
\author[2]{P.V.R.G. Silva\footnote{precchia@ifi.unicamp.br}}
\affil[1]{\small Instituto de Matem\'atica, Estat\'{\i}stica e Computa\c c\~ao Cient\'{\i}fica - UNICAMP \\
13083-859 Campinas, SP, Brazil}
\affil[2]{\small Instituto de F\'{\i}sica Gleb Wataghin, Universidade Estadual de Campinas - UNICAMP \\
13083-859 Campinas, SP, Brazil}

\maketitle

\begin{abstract}
In high-energy hadron-hadron collisions, the dependence of the total cross-section
($\sigma_{tot}$) with the energy still constitutes an open problem for QCD.
Phenomenological analyses
usually relies on analytic parameterizations provided by the Regge-Gribov 
formalism and fits to the experimental data. 
In this framework, the singularities
of the scattering amplitude in the complex angular momentum plane determine
the asymptotic behavior of $\sigma_{tot}$ in terms of the energy.
Usual applications connect simple and triple pole singularities with asymptotic power and
logarithmic-squared functions of the energy, respectively. More restrict applications have considered
as a leading component for $\sigma_{tot}$ an \textit{empirical} function consisting of
a logarithmic raised to a \textit{real} exponent, which is treated as a free fit parameter. 
With this function, data reductions lead to good descriptions of the experimental data
and \textit{real} (\textit{not integer}) values of the exponent.
In this paper, making use of two independent formalisms (fractional calculus and Mellin transform),
we first show that the singularity associated with this empirical function is a branch point and
then we explore the mathematical consequences of the result and possible physical 
interpretations.
After reviewing the determination of the singularities from asymptotic forms of interest 
through the Mellin transform, we demonstrate that the same analytic results can be obtained by means of
the Caputo fractional derivative, leading, therefore,  to fractional calculus interpretations. 
Besides correlating Mellin 
transform, \textit{non-local} fractional derivatives and exploring
the generalization from integer to real exponents and derivative orders, this fractional calculus
result may provide insights for physical interpretations on the asymptotic
rise of $\sigma_{tot}$. We also illustrate the applicability of the parameterizations by means of novel
fits to $\sigma_{tot}$ data.
\end{abstract}



\newpage

\vspace{0.5cm}

\textbf{Contents}

\vspace{0.2cm}

1. Introduction

2. Analytic Parametrization and Mellin Transform

\ \ \ 2.1 Analytic Parametrization and a Canonical Function

\ \ \ 2.2 Mellin Transform and Singularities

3. Fractional Calculus and the Caputo Derivative

\ \ \ 3.1 Introduction and Motivation

\ \ \ 3.2 Fractional Calculus

\ \ \ 3.3 Fractional Integral Operator

\ \ \ \ \ \ \ \ 3.3.1 Fractional Integral in the Riemann-Liouville Sense

\ \ \ 3.4 Fractional Derivative Operators

\ \ \ \ \ \ \ \ 3.4.1 Fractional Derivative in the Riemann-Liouville Sense

\ \ \ \ \ \ \ \ 3.4.2 Fractional Derivative in the Caputo Sense

\ \ \ 3.5 Connecting Asymptotic Forms and Singularities

4. Discussion

5. Summary, Conclusions and Final Remarks

\vspace{0.2cm}

Appendix A Total Cross Section and the Regge-Gribov Formalism

\ \ \ \ \ \ \ \ \ \ \ \ \ \ \ \ \ A.1 Total Hadronic Cross Section

\ \ \ \ \ \ \ \ \ \ \ \ \ \ \ \ \ A.2 Basic Phenomenological Concepts

\ \ \ \ \ \ \ \ \ \ \ \ \ \ \ \ \ A.3 Analytic Parameterizations 

Appendix B Data Reductions to $pp$ and $\bar{p}p$ Total Cross-Sections

\ \ \ \ \ \ \ \ \ \ \ \ \ \ \ \ \ B.1 Fit Procedures and Results

\ \ \ \ \ \ \ \ \ \ \ \ \ \ \ \ \ B.2 Discussion on the Fit Results



\section{Introduction}
\label{sintro}

High-Energy Physics  deals with the study of the inner structure of matter, 
the sub-nuclear particles and their interactions \cite{halzen}. The main experimental tool consists of
collision of these particles (protons, pions, kaons,...) at high energies, namely center-of-mass energies
above 10 $m_p \sim$ 10 GeV, where $m_p$ is the proton mass.
Presently, the highest energies reached in accelerators, with particles and anti-particles,
concern proton-proton ($pp$) and antiproton-proton ($\bar{p}p$) collisions
at 8 TeV and $\sim$ 2 TeV, respectively. These strong (hadronic) interactions are expected to be
described by the Quantum Chromodynamics (QCD), a non-Abelian gauge field theory \cite{halzen,qcd}.

Among the physical observables characterizing a particle collision process, the total cross
section ($\sigma_{tot}$) plays a central role. On the one hand, beyond being physically related with
the probability and also an effective area of interaction, a large amount of experimental
data, as a function of the collision energy, has been obtained from several
scattering processes, providing solid information on the empirical behavior of this quantity.
On the other hand, however, the theoretical description of its dependence on the energy
constitute a long standing challenge for QCD.
Indeed, the total cross-section is obtained from the imaginary part of the forward
\textit{elastic} scattering amplitude $\mathcal{A}(s,t)$ through the optical theorem, which at high energies reads \cite{pred}
\begin{eqnarray}
\sigma_{\mathrm{tot}}(s) = \frac{\mathrm{Im}\,\mathcal{A}(s,t=0)}{s},
\label{ot}
\end{eqnarray}
where $s$ and $t$ are the energy and momentum transfer squared in the center of mass system
(Mandestam variables) and  $t=0$ means the forward direction.
Therefore, the determination of $\sigma_{tot}(s)$ demands
a theoretical result for the forward \textit{elastic} amplitude in terms of the energy,
valid in all region above the physical threshold. However, as a soft scattering
state (large distances and small momentum transfer), the perturbative QCD techniques
cannot be applied due to the increase of the coupling constant as the momentum transfer
decreases, namely the dynamics involved in the
\textit{elastic scattering} is intrinsically 
nonperturbative.  Presently, the crucial point concerns
the lack of a nonperturbative framework able to provide \textit{from the first principles} of QCD
a description of $\sigma_{tot}(s)$ for all $s$.

A historical and formal result on the rise of $\sigma_{tot}(s)$, at the asymptotic energy region 
($s \rightarrow \infty$), is the upper bound derived by Froissart and Martin (FM)
\cite{froissart,martin1,martin2,lukamartin}
\begin{eqnarray}
\sigma_{tot}(s) < c \ln^2(s/s_0),
\label{fmb}
\end{eqnarray}
where $s_0$ is an energy scale and $c$ a constant.
More recently, 2011, and yet in a formal
(theoretical) context, the possibility of a rise of $\sigma_{tot}(s)$
faster than the above log-squared bound, without violating unitarity, was discussed by
Azimov \cite{azimov}.

Beyond specific phenomenological models \cite{igor,giulia}, the behavior of $\sigma_{tot}(s)$ is
usually investigated by means of \textit{forward amplitude analyses}, consisting of tests
of different 
analytic parameterizations for $\sigma_{tot}(s)$ and the $\rho$ parameter 
(the ratio between the real and imaginary parts of the forward amplitude) and fits
to the experimental data.
The analytic parameterizations are constructed on the bases of $S$-Matrix and Regge-Gribov 
formalism \cite{pred,land,collins}. In this context, the
singularities of the scattering amplitude in the complex angular momentum $J$-plane
($t$-channel) determine the asymptotic energy dependence of the total cross-section ($s$-channel).
As we shall discuss, simple, double and triple pole singularities result in power, logarithmic
and logarithmic-squared laws, respectively, for $\sigma_{tot}(s)$. The standard picture indicates
a leading log-squared dependence at the highest energies \cite{compete1,compete2,pdg14,pdg16},
in accordance with the FM
bound, Eq. (\ref{fmb}) and associated, therefore,
with a triple pole singularity in the complex $J$-plane.

Beyond this log-squared, $\ln^{2}(s/s_0)$ (L2), leading dependence, other amplitude analyses have considered
an \textit{empirical} leading log-raised-to-gamma form, $\ln^{\gamma}(s/s_0)$ (L$\gamma$), with $\gamma$ as a real free fit parameter,
a parametrization introduced by Amaldi et al. in 1977 \cite{amaldi}.
Besides providing equivalent descriptions of the experimental data
available \cite{amaldi,ua42,velasco}, including the recent LHC data \cite{fms1,fms2,ms1,ms2,fms17a,fms17b},
this empirical parametrization is also useful as a quantitative check on the behavior of $\sigma_{tot}(s)$
(dictated by the experimental data) in respect the FM bound, namely how close to 2 the
extracted $\gamma$-value lies: equal 2, below or above 2.

In this respect, although based on different approaches and datasets,
the analyses quoted in the previous paragraph, suggest $\gamma$-values 
exceeding 2 (typically $2.2 - 2.3$ in \cite{ms1,ms2}), which contrasts with the result obtained by the COMPAS Group,
favoring a $\gamma$-value below 2: $1.98 \pm 0.01$ \cite{pdg14, pdg16}. To our purposes, 
the main point to notice is: 
once amplitude analyses with the exponent $\gamma$ 
treated as a free fit parameter have favored \textit{real} (\textit{not integer}) \textit{and positive} $\gamma$ \textit{values}, 
the associated singularity may not be a triple pole.

In spite of the aforementioned useful properties,
the L$\gamma$ law has had only an \textit{empirical} basis, without further justification or 
devised connections with a formal approach and that is the point we are interested in here,
at least in a \textit{mathematical} context.
Taking account of the absence of a pure nonperturbative QCD description of $\sigma_{tot}(s)$,
we explore some mathematical aspects related to the L$\gamma$ law, looking
for possible connections with the Regge-Gribov formalism. 
With that in mind, we shall consider two independent frameworks: the Fractional Calculus
(which is a generalization of the classical integer order Calculus to
real or complex order) \cite{fc1,fc2,fc3}
and the Mellin Integral Transform (which can connect asymptotic
forms of real valued functions of real argument, with the singularities of the
transformed function in a complex plane) \cite{bateman,bbo,joubert}.

Our specific goals are: (1) in a kind of inverse approach, to review the determination
of the singularities associated with the asymptotic analytic components of $\sigma_{tot}(s)$
by means of the Mellin transform, showing, in particular, that the L$\gamma$ law,
with $\gamma > 0$, real and not integer,
can be associated with a branch point singularity; (2) to demonstrate that the same analytic results can be
obtained in a fractional calculus approach, through the Caputo fractional
derivative \cite{cfd1,cfd2}; (3) to explore the possible connections among the Mellin transform, the
\textit{nonlocal} character of the Caputo fractional derivative (memory effects) 
and the Regge-Gribov formalism.
In addition, we also illustrate the applicability of the full parametrization (including
lower energy components), by developing
novel fits to all the $\sigma_{tot}$ data presently available
from $pp$ and
$\bar{p}p$ scattering in the energy region 5 GeV - 8 TeV, which lead to real
(not integer) $\gamma$ value exceeding 2: $\gamma \approx 2.2 \pm 0.1$.

The paper is organized as follows. In Sect. 2, after identifying a characteristic function in the
analytic parameterizations for $\sigma_{tot}(s)$, 
we review the use of the Mellin transform
in the determination of the associated singularities. In Sect. 3, after recalling some basic concepts 
on Fractional Calculus,
we develop the novel calculations connecting the characteristic function with the corresponding
singularities. Sect. 4 is devoted to a discussion on all the obtained results and
possible physical interpretations associated. Our conclusions and
final remarks are the contents of Sect. 5. 
In Appendix A, we refer to the $\sigma_{tot}$ concept and the experimental data
presently available at the highest energies, presenting also a short review
on the construction of the analytic parametrization
for $\sigma_{tot}(s)$ in the Regge-Gribov context. In Appendix B, we
illustrate the applicability of the parameterizations by means of novel fits
to $\sigma_{tot}$ data from $pp$ and $\bar{p}p$ scattering.

\section{Analytic Parametrization and Mellin Transform}
\label{smellin}

After identifying a characteristic function of the energy, in the 
asymptotic parameterizations
for the total cross section, we review the determination of the
corresponding singularities in the complex $J$-plane by means of
the Mellin transform; see, for example,
\cite{pred} (Appendix B ), \cite{fr} (Sect. 2.8), \cite{smatrix} (Sect. 3.6).

\subsection{Analytic Parametrization and a Canonical Function}

As reviewed in Appendix A (and quoted references), the analytic parametrization
for the total cross section, introduced by Amaldi et al. \cite{amaldi} and used
in several analyses \cite{ua42,velasco,fms1,fms2,ms1,ms2,fms17a,fms17b}, can be expressed by

\begin{eqnarray}
\sigma_{tot}(s) = a_1 [\tilde{s}]^{-b_1} + \tau a_2 [\tilde{s}]^{-b_2} +
A + B \ln^{\gamma} ( \tilde{s} ), 
\label{general}
\end{eqnarray}
where $a_i, b_i, i = 1,2$, $A$, $B$, $\gamma$ are  real parameters, $\tau = -1$ for $pp$ and 
$\tau = +1$ for $\bar{p}p$ and
\begin{eqnarray}
\tilde{s} = \frac{s}{s_0},
\label{scale}
\end{eqnarray}
with $s_0$ an energy scale.

For our purposes it is important to note \cite{fr,smatrix,edenbook} that all the components in Eq. (\ref{general}) are particular
cases of a \textit{canonical} (dimensionless) function given by
\begin{eqnarray}
f_{ca}(\tilde{s}) = \tilde{s}^{\alpha} \ln^{\gamma}(\tilde{s}),
\label{canonical}
\end{eqnarray}
where, for short, we denote here $\alpha_0 - 1 \equiv \alpha$, with $\alpha_0$ the intercept
of the trajectory (see Appendix A.2). 
Indeed, for $\gamma = 0$ and different $\alpha$ values, we have all the power (and constant) laws 
in Eq. (\ref{general})
associated with
Reggeons and Pomerons. For $\alpha = 0$ and $\gamma = 1, 2$
the logarithmic dependences representing the Pomeron are obtained (Appendix B). In case of $\gamma$ a real (not integer)
parameter ($\alpha = 0$ for the L$\gamma$ law), the association with a cross section demands real values for the logarithmic function and therefore we
have an additional condition:
\begin{eqnarray}
\tilde{s} \ge 1 \quad \mathrm{for\ real\ (not\ integer)}\ \gamma
\nonumber
\end{eqnarray}

In the next Subsection we shall obtain the singularities associated with the
asymptotic canonical function Eq. (\ref{canonical}) by means of the Mellin transform.

\subsection{Mellin Transform and Singularities}

The Mellin transform connects a real valued function $\phi(t)$, defined on the real axis 
$t \in (0, \infty)$, with a function $\Phi(z)$, defined in the complex
plane $z = x + i y$, by the relation \cite{bateman,bbo}:

\begin{eqnarray}
\Phi(z) = \int_{0}^{\infty} t^{z - 1} \phi(t) dt \equiv \mathcal{M}[\phi(t), z]. 
\label{mellin}
\end{eqnarray}
The integral exists in the region $a < x < b$, named strip of definition
(or fundamental strip),
where $a$ and $b$ depend on $\phi(t)$. 

For our purposes, the main ingredient here
is the correspondence between the asymptotic expansion of the function
$\phi(t)$ and the singularities of the transformed function $\Phi(z)$ in the
complex $z$-plane \cite{bbo,joubert}. 
To this end, from the definition, it is straightforward
to show that $\Phi(-z) = \mathcal{M}[\phi(1/t), z]$, which allows to change the
kernel to $t^{-z -1}$, useful for identifying the singularity on the $z$-plane.
Our aim is to use as input the canonical function Eq. (\ref{canonical}), obtaining the
associated singularities.

Firstly, by changing the variable $t = 1/ \tilde{s}$ in Eq. (\ref{mellin}) and
for $\Phi(-z) \equiv F(J)$ (with $J$ the complex angular momentum), we obtain

\begin{eqnarray}
F(J) =  \int_{0}^{\infty} \tilde{s}^{-J - 1} \phi(1/\tilde{s}) d\tilde{s}. 
\label{}
\end{eqnarray}

In order to treat the general case, where $\gamma$ can take on real (not integer) values,
we connect $\phi(1/\tilde{s})$ with the canonical function, Eq. (\ref{canonical}), by means
of a Heaviside function (note the inversion in the independent variables),

\begin{eqnarray}
\phi(1/\tilde{s}) \equiv \theta(\tilde{s} - 1) f_{ca}(\tilde{s}),
\nonumber
\end{eqnarray}
leading to
\begin{eqnarray}
F(J) = \int_{1}^{\infty} \tilde{s}^{\alpha - J - 1} \ln^{\gamma} (\tilde{s}) d\tilde{s}.
\end{eqnarray}

At last, with a change of variable, $\tilde{s} = 1/\xi$, the above integral reads
\begin{eqnarray}
F(J) = \int_{0}^{1} \xi^{(J - \alpha) - 1} \ln^{\gamma}\left(\frac{1}{\xi}\right) d\xi
\end{eqnarray}
and from \cite{grads} (page 550, formula 4.272.6), we obtain the final result

\begin{eqnarray}
F(J) = \frac{\Gamma(\gamma + 1)}{(J - \alpha)^{\gamma + 1}},
\label{singmellin}
\end{eqnarray}
for $\gamma > -1$, Re$[J - \alpha] > 0$  and where $\Gamma$ is the
Euler gamma function.

We see that, for $\gamma = n = 0, 1, 2$, the corresponding singularities at
$J = \alpha$ are poles of order $n+1$ and for real (not integer) $\gamma > 0$ values,
the associated singularities are branch points. 
In the next Section, we shall correlate the results given by Eqs. (\ref{canonical}) and (\ref{singmellin}) in a completely
different mathematical context, which shall provide further interesting interpretations.

\section{Fractional Calculus and the Caputo Derivative}
\label{sfc}

\subsection{Introduction and Motivation}

As we have mentioned (see Appendix A and quoted references), in the Regge-Gribov formalism,
a simple pole in the complex angular momentum
plane is associated with a power law for the total cross section as a function of the energy:
\begin{eqnarray}
\frac{1}{J - \alpha}
\qquad \Longleftrightarrow \qquad
\tilde{s}^{\alpha}.
\nonumber
\end{eqnarray}
Note that these are particular cases of Eqs. (\ref{canonical}) and (\ref{singmellin})
for $\gamma = 0$. 

As discussed in \cite{land} (Sect. 2.3), 
higher order
poles can be generated by derivatives of the simple pole,
\begin{eqnarray}
\frac{d^n}{d \alpha^n}\left(\frac{1}{J - \alpha}\right) = 
\frac{n!}{(J - \alpha)^{n+1}},
\quad
n = 1, 2,...,
\label{poles}
\end{eqnarray}
with $n +1 \equiv N$ the order of the pole. Translating this derivative
to the power-law, we obtain
\begin{eqnarray}
\frac{d^n}{d \alpha^n} \tilde{s}^{\alpha} = \tilde{s}^{\alpha} \ln^n (\tilde{s}),
\quad
n = 1, 2,...
\label{logn}
\end{eqnarray}

We notice that Eqs. (\ref{poles}) and (\ref{logn}) correspond to particular cases of
Eqs. (\ref{canonical}) and (\ref{singmellin}) for poles singularities and/or
integer powers in the logarithmic functions. Moreover, from Eqs. (\ref{poles}) and (\ref{logn}),
this ``integer"\ character
is associated with the integer order of the derivatives respect to $\alpha$.
It seems, therefore, natural to look for a generalization of this result
with bases on extensions of integer to real orders, namely the Fractional Calculus \cite{fc1,fc2,fc3}.

In this section, after reviewing some basic definitions of Fractional Calculus
and introducing the fractional derivative in the Caputo sense \cite{cfd1,cfd2}, we generalize
Eqs. (\ref{poles}) and (\ref{logn}) for a real order (not integer) derivative,
obtaining Eqs. (\ref{canonical}) and (\ref{singmellin}). Discussions on the
implications of this result and possible physical interpretations are presented
in Sect. 4.

\subsection{Fractional Calculus}

In 1695, after a letter by l'H\^opital to Leibniz, where the former questioned the meaning of a semi-derivative of $y(x)$ (dependent variable) in relation to $x$
(independent variable), fractional calculus was born \cite{MillerRoss}. 
Although historically associated to a semi-derivative (fractional order), the true name fractional
calculus concerns calculus of arbitrary order.
The first attempt to unify several areas involved with the fractional calculus, took place in the first 
International Congress, realized at New Haven, in 1974 \cite{Ross1974}. The fractional calculus has 
then grown and several definitions involving the integral 
operator and derivative operators have appeared in the literature \cite{capelastenreiro}. 

We note that there are more than one way to introduce the concept of 
fractional derivatives, one of them, by means of the corresponding integral operator with a singular kernel \cite{Diethelm,Mainardi1} and, in some papers, 
by means of a convenient limit process \cite{katugampola,khalil} and more recently through an integral with a nonsingular kernel \cite{CaputoF,losadanieto}. 
On the other hand, in a recent and interesting paper \cite{ortigueiramachado} the authors discuss a criterion that an operator must satisfy to be interpreted 
as a fractional derivative operator, similar to the one proposed by Ross \cite{RossCriteria}. 

Here, we will employ the first approach, i.e., using the fractional 
integral operator, we introduce the fractional differential operator. To this end, we first treat the fractional integral operator and after that, the fractional 
derivative operator, in the Caputo sense, only.

\subsection{Fractional Integral Operator}
There are several ways to introduce a fractional operator of arbitrary order, we mention two of them, i.e., in the Riemann-Liouville sense and their 
variations \cite{Kilbas-Srivastava-Trujillo.2006A} and in the Hadamard sense \cite{daniela}, each one defined by a particular class of functions where 
the operators can be applied. We will only consider spaces where the functions are continuous or continuous by parts \cite{SamkoKilbasMarichev1993}. Thus, 
we introduce the Riemann-Liouville fractional integral and the Caputo fractional derivative, defined in terms of the Riemann-Liouville fractional integral.

\subsubsection{Fractional Integral in the Riemann-Liouville Sense} Let $\Omega=[a, b]$ with $- \infty < a < b < \infty$ be a finite interval on the real axis $\mathbb{R}$. 
The integral of arbitrary order in the Riemann-Liouville sense, of order $\mu \in \mathbb{C}$, with $\mbox{Re}(\mu)>0$, denoted by, $(J_{a+}^\mu f)(x) 
\equiv {}_aJ_x^\mu f(x)$ is defined by:

\begin{equation}
\label{2156}
(J_{a+}^\mu f)(x):=\frac{1}{\Gamma(\mu)}\int_{a}^{x}\frac{f(t)dt}{(x-t)^{1-\mu}},
\qquad x>a,  \qquad \mbox{Re}(\mu)>0.  
\end{equation}
This integral is the so-called fractional integral in the Riemann-Liouville sense on the left. We can introduce another one on the 
right \cite{Kilbas-Srivastava-Trujillo.2006A}. The integral of arbitrary order in the Riemann-Liouville sense, can be extended on the real axis $\mathbb{R}$. 

\subsection{Fractional Derivative Operators}
As we already said, we are interested in the fractional derivative in the Caputo sense, but as this definition can be done in terms of the fractional 
derivative in the Riemann-Liouville sense, we first introduce the fractional derivative in the Riemann-Liouville sense by means of the Riemann-Liouville integral.

\subsubsection{Fractional Derivative in the Riemann-Liouville Sense} The Riemann-Liouville derivative in a finite interval on the real axis, of order $\mu \in 
\mathbb{C}$ with $\mbox{Re}(\mu) 
\geq 0$, denoted by, ${\mbox{D}}_{a+}^\mu y$ is defined by 
$$
({\mbox{D}}_{a+}^\mu y)(x):=\left(\frac{d}{dx}\right)^n\left(J_{a+}^{n-\mu}y
\right) (x)
$$
with $ n = [\mbox{Re}(\mu)]+ 1;\,\, x > a$, where $[\mbox{Re}(\mu)]$ means the integer part of $\mbox{Re}(\mu)$. 

Also, this derivative is named fractional derivative in the Riemann-Liouville sense on the left. 
Analogously, we can define another one on the 
right \cite{Kilbas-Srivastava-Trujillo.2006A}. The derivative of an arbitrary order in the Riemann-Liouville sense, can be extended on the real axis $\mathbb{R}$. 
In words, the fractional derivative in the Riemann-Liouville sense, is equal to the derivative of integer order of a fractional integral.

\subsubsection{Fractional Derivative in the Caputo Sense}
Let $[a,b]$ be a finite interval on the real axis $\mathbb{R}$ and ${\mbox{D}}_{a+}^{\mu}[y(t)](x) \equiv ({\mbox{D}}_{a+}^{\mu} y)(x)$ be the derivative 
in the  Riemann-Liouville sense, with $\mu \in \mathbb{C}$ and $\mbox{Re}(\mu) \geq 0$. The fractional derivative $({}^C{\mbox{D}}_{a+}^{\mu} y)(x)$ of 
order $\mu \in \mathbb{C}$ with $\mbox{Re}(\mu) \geq 0$ on $[a,b]$ is defined by means of the Riemann-Liouville 
derivative as follows
\begin{equation}
\label{2264}
({}^C{\mbox{D}}_{a+}^{\mu} y)(x):=\left({\mbox{D}}_{a+}^{\mu}\left[y(t)-\sum_{k=0}^{n-1}
\frac{y^{(k)}(a)}{k!}(t-a)^k\right]\right)(x)
\end{equation}
 with 
\begin{equation}
\label{2266}
n= [\mbox{Re}(\mu)]+1 \quad \mbox{for} \quad \mu \notin \mathbb{N} \quad {\mbox{and}} \quad 
n=\mu \quad \mbox{for} \quad \mu \in \mathbb{N}.
\end{equation}
This derivative is named \textit{Caputo fractional derivative} of order $\mu$ on the left. 
The Caputo derivative $({}^C{\mbox{D}}_{a+}^{\mu} y)(x)$ is defined for functions $y(x)$ such that the Riemann-Liouville derivative 
on the left Eq. (\ref{2264}) exists. Particularly, they are defined for $y(x)$ in the space of functions $AC[a, b]$ absolutely continuous. 
The following theorem is valid \cite{Kilbas-Srivastava-Trujillo.2006A}.

\noindent {\sf Theorem. Caputo fractional derivative} \label{def:84} 

Let $\mbox{Re}(\mu) \geq 0$ and $n$ given by Eq. (\ref{2266}). If $y(x) \in AC^{n}[a, b]$, then the Caputo derivative 
$({}^C{\mbox{D}}_{a+}^{\mu} y)(x)$ exists in almost all points of $[a, b].$

\noindent (a) If $\mu \notin \mathbb{N}, ({}^C{\mbox{D}}_{a+}^{\mu} y)(x)$ is represented by
\begin{equation}
\label{caputo}
({}^C{\mbox{D}}_{a+}^{\mu}y)(x) := \frac{1}{\Gamma(n-\mu)}\int_{a}^{x}
\frac{y^{(n)}(t)dt}{\left(x-t\right)^{\mu-n+1}}=:(J_{a+}^{n-\mu}{\mbox{D}}^{n}y)(x)  
\end{equation}
with ${\mbox{D}}=d/dx$ and $n=[\mbox{Re}(\mu)]+1.$

\noindent (b) If $\mu = n \in \mathbb{N}_{0}$, then $({}^C{\mbox{D}}_{a+}^{n}y)(x)$ is represented by 
\begin{equation}
\label{2282}
({}^C{\mbox{D}}_{a+}^{n}y)(x):= \frac{d^n}{dx^n}y(x).
\end{equation}
Specifically, we have $({}^C{\mbox{D}}_{a+}^{0}y)(x):=y(x)$, which recovers the function. For the proof, see \cite{Kilbas-Srivastava-Trujillo.2006A}.

Contrary to the Riemann-Liouville derivative, the fractional derivatives in the Caputo sense, 
are equal to the fractional integral of the derivative of integer order.

\subsection{Connecting Asymptotic Forms and Singularities}

In this Section we shall generalize the integer order results expressed by Eqs. (\ref{poles}) and (\ref{logn})
to real order by using the \textit{Caputo fractional derivative}, Eq. (\ref{caputo}), here denoted by 
\begin{eqnarray}
\frac{d^{\mu}}{dx^{\mu}}
\end{eqnarray}
with $\mu >0$ a real order.

Following the steps in Sect. 3.1 (integer order),
let us first evaluate the Caputo derivative of a simple pole, $y(x) = (x_0 - x)^{-1}$,
with $x_0$ fixed.
From Eq. (\ref{caputo}) and taking the limit $a \rightarrow - \infty$,
\begin{eqnarray}
\frac{d^{\mu}}{dx^{\mu}} (x_0-x)^{-1} &=&
\frac{1}{\Gamma(n-\mu)}\int_{- \infty}^{x}
\frac{1}{(x-t)^{\mu+1-n}}
\frac{d^n}{dt^n} (x_0 - t)^{-1} dt \nonumber \\ 
&=& \frac{n!}{\Gamma(n-\mu)}\int_{- \infty}^{x}
\frac{dt}{(x_0-t)^{n+1} (x-t)^{\mu+1-n}}, \quad n- \mu > 0.
\nonumber
\end{eqnarray}
By changing the variable, $x-t=\xi$,
\begin{eqnarray}
\frac{d^{\mu}}{dx^{\mu}} (x_0-x)^{-1} = 
 \frac{n!}{\Gamma(n-\mu)}\int_{0}^{\infty}
\frac{\xi^{n-\mu-1}}{(x_0-x+\xi)^{n+1}} d\xi.
\label{int}
\end{eqnarray}
From \cite{grads}, page 285 and rearranging terms,
\begin{eqnarray}
\int_{0}^{\infty} \frac{x^{\lambda-1}}{(x + 1/\beta)^{n+1}} dx =
(-1)^n \pi \beta^{n+1-\lambda} 
{\lambda - 1 \choose n}
\frac{1}{\sin \pi \lambda},
\nonumber
\end{eqnarray}
with $|\mathrm{arg}\, \beta| < \pi$ and $0 < \mathrm{Re}\, \lambda < n+1$.
By comparing with Eq. (\ref{int}) we have

\begin{eqnarray}
\frac{d^{\mu}}{dx^{\mu}} (x_0-x)^{-1} =
\frac{n!}{\Gamma(n-\mu)} (-1)^n \pi 
\left(\frac{1}{x_0-x}\right)^{1+\mu}
{n - \mu - 1 \choose n} \frac{1}{\sin \pi (n-\mu)}.
\nonumber
\end{eqnarray}

Now, from $\pi/\sin (\pi z) = \Gamma(z)\Gamma(1-z)$, with $z=-\mu$,
we obtain the result

\begin{eqnarray}
\frac{d^{\mu}}{dx^{\mu}} (x_0-x)^{-1} =
\frac{\Gamma(\mu + 1)}{(x_0 - x)^{\mu+1}}.
\label{capbranch}
\end{eqnarray}

Let us now evaluate the Caputo derivative of a power function,
$y(x) = b^x$, with $b$ a real constant. From Eq. (\ref{caputo}),

\begin{eqnarray}
\frac{d^{\mu}}{dx^{\mu}} b^x &=&
\frac{1}{\Gamma(n-\mu)}\int_{a}^{x}
\frac{1}{(x-t)^{\mu+1-n}}
\frac{d^n}{dt^n} b^t dt \nonumber  \\ 
&=& \frac{1}{\Gamma(n-\mu)}\int_{a}^{x}
\frac{b^t \ln^n(b)}{(x-t)^{\mu+1-n}} dt \quad n- \mu > 0.
\nonumber
\end{eqnarray}
By changing the variable, $x-t=\lambda$,

\begin{eqnarray}
\frac{d^{\mu}}{dx^{\mu}} b^x =
\frac{b^x \ln^n(b)}{\Gamma(n-\mu)}\int_{0}^{x-a}
b^{-\lambda} \lambda^{n-\mu-1} d\lambda.
\nonumber
\end{eqnarray}
For $b^{-\lambda} = e^{\xi}$, the above integral can be put in the form,

\begin{eqnarray}
\frac{(-1)^{n-\mu}}{\ln^{n-\mu}(b)}\int_{0}^{(a-x)\ln(b)}
e^{\xi} \xi^{n-\mu-1} d\xi,
\nonumber
\end{eqnarray}
so that

\begin{eqnarray}
\frac{d^{\mu}}{dx^{\mu}} b^x &=&
\frac{b^x}{\Gamma(n-\mu)}\ln^{\mu}(b) (-1)^{n-\mu}
\int_{0}^{(a-x)\ln(b)}
e^{\xi}\, \xi^{n-\mu-1} d\xi \nonumber  \\ 
&=& \frac{b^x}{\Gamma(n-\mu)} \ln^{\mu}(b)\gamma(n-\mu,[x-a]\ln b),
\nonumber
\end{eqnarray}
where,
\begin{eqnarray}
 \gamma(a,x) = \int_{0}^{x} t^{a-1} e^{-t} dt
\nonumber
\end{eqnarray}
is the incomplete gamma function.

Now, for $a \rightarrow - \infty$, we obtain
$\gamma(n-\mu, \infty) = \Gamma(n-\mu)$ and therefore the result

\begin{eqnarray}
\frac{d^{\mu}}{dx^{\mu}} b^x = b^x \ln^{\mu}(b).
\label{caplogr}
\end{eqnarray}

At last, from Eqs. (\ref{capbranch}) and (\ref{caplogr}) and translating to the notation in Sect. 2,
namely
\begin{eqnarray}
\mu \rightarrow \gamma,\quad
x \rightarrow \alpha, \quad
x_0 \rightarrow J, \quad
b \rightarrow \tilde{s},
\nonumber
\end{eqnarray}
we obtain our main results in the form

\begin{eqnarray}
\frac{d^{\gamma}}{d\alpha^{\gamma}} \left(\frac{1}{J - \alpha}\right) =
\frac{\Gamma(\gamma + 1)}{(J - \alpha)^{\gamma+1}}
\label{branch}
\end{eqnarray}
and
\begin{eqnarray}
\frac{d^{\gamma}}{d\alpha^{\gamma}} \tilde{s}^{\alpha} = \tilde{s}^{\alpha} \ln^{\gamma}(\tilde{s}).
\label{logr}
\end{eqnarray}
which generalize Eqs. (\ref{poles}) and (\ref{logn}), for real (not integer) $\gamma$, corresponding
to the functions $F(J)$ and $f_{ca}(\tilde{s})$,
Eqs. (\ref{canonical}) and (\ref{singmellin}), respectively, treated in Sect. 2 by means of the
Mellin transform.

It is interesting to note that the above generalization of the integer order results,
Eqs. (\ref{poles}) and (\ref{logn}), to real order results, Eqs. (\ref{branch}) and (\ref{logr})
(or (\ref{canonical}) and (\ref{singmellin})), can be obtained through a simple and compact
``prescription", by associating
\begin{eqnarray}
\mathrm{integer}\ n \quad &\Rightarrow& \quad \mathrm{real}\ \gamma \nonumber \\
n! = \Gamma(n + 1) \quad &\Rightarrow& \quad \Gamma(\gamma + 1) \nonumber 
\end{eqnarray}
which constitute the basics of the Fractional Calculus. However, note that Eqs. (\ref{branch}) and (\ref{logr})
were deduced through a specific fractional derivative (Caputo sense) and specific lower
integration limit ($a \rightarrow - \infty$ in Eq. (\ref{caputo})).

\section{Discussion}
\label{sdisc}

In Sects. 2 and 3 we have shown that through both Mellin integral transform
and Caputo fractional derivative, the L$\gamma$ law with asymptotic
dependence $\ln^{\gamma}(\tilde{s})$, for real (not integer) $\gamma >$ 0, can be associated with a singularity 
consisting of a branch point at $J = \alpha$ and given by 
\begin{eqnarray}
\frac{1}{(J - \alpha)^{\gamma + 1}}.
\label{bp}
\end{eqnarray}
This corresponds to a strictly mathematical  result.

In looking for consequences an possible physical interpretations, we first note that
in the Regge-Gribov formalism, branch points (Regge cuts) can be related to the
exchange of not one, but two or more Reggeons (or Pomerons). The framework (Gribov
calculus) is based on perturbative techniques and a class of Feynmann diagrams, involving
both Mandesltam variables $s$ and $t$. The \textit{asymptotic} amplitude for $N$
identical Reggeon exchanges is given by
(see, for example, \cite{collins}, Sect. 8.5 and \cite{pred}, Sect. 5.10)
\begin{eqnarray}
\mathcal{A}_{NR}(s,t) \sim \frac{\tilde{s}^{\alpha_c (t)}}{\ln^{N-1}(\tilde{s})},
\nonumber
\end{eqnarray}
corresponding to a branch point (attached cut) located at $J = \alpha_c(t)$.
Since $N \geq 2$, we see that, at $t=0$, the effect of the multiple exchange is
to tame the power dependence of the total cross section. It is expected that these
contributions may be important at large $t$-values  (where the
perturbative approach does apply), but not in the nonperturbative region
of small $t$, in special as $t \rightarrow 0$, namely the total 
cross-section\footnote{The exchange of two Pomerons related, however, to two
simple pole singularities is discussed in \cite{cudell1} and \cite{cudell2}.}.

In what concerns the asymptotic L$\gamma$ law, since the phenomenological analyses  
indicate $\gamma > 0$ and real (not integer) values, the corresponding
branch point we have obtained, Eq.  (\ref{bp}), cannot be associated with multiple
exchanges of Reggeons (or Pomerons).
On the other hand, following the physical meaning of the integer order results, Eqs. (\ref{poles}) and
(\ref{logn}), it seems reasonable to interpret this leading component of the $\sigma_{tot}(s)$
as associated with the branch point (\ref{bp}) in the
complex $J$-plane and located at $J=\alpha$ and $\alpha=0$ (intercept $\alpha_0 = 1$). If that is the case,
the physical origin of this singularity might be related to some nonperturbative
effect which, however, seems not easy to be identify.

Nonetheless, two aspects involved in the mathematical result suggest
some interesting interpretations in the fractional calculus context. 
Firstly, the Regge-Gribov formalism is
essentially based on simple poles (the Regge poles) and not
double or triple poles, which constitute mathematical possibilities
(Sect. 2.3 in \cite{land}). Second, the branch point in Eq. (\ref{branch}) is
a result of the Caputo fractional derivative applied just to a simple
pole, having, therefore, properties distinct from those obtained through
derivatives of integer order (double pole, triple pole, ...).
Indeed, once expressed in terms of an integral, the Caputo
derivative, Eq. (\ref{caputo}), has an intrinsic nonlocal character.
Specifically,
the Caputo fractional derivative of a given function does not depend
only on the point where it is evaluated (as in the integer case),
but also on the neighboring points and, moreover, on the values of the function
between that point and the lower integration limit
(in this case, $a \rightarrow - \infty$). 
In this context, the fractional calculus
results, Eqs. (\ref{branch}) and (\ref{logr}), embody more
information on the process investigated than the integer order results, 
Eqs. (\ref{poles}) and (\ref{logn}), as nonlocality and memory effects.
In what concerns singularities, not integer $\gamma$-values (branch point)
might be related to interpolations between integer values (poles).
At last, the aforementioned nonlocal character (or additional
information), provided by the fractional
derivative, may points toward interpretations on the presence of some
unknown nonperturbative effects.

\section{Summary, Conclusions and Final Remarks}
\label{sconclu}

As commented in our introduction, once the total cross section is directly connected
with the forward elastic scattering (optical theorem, Eq. (\ref{ot})), its theoretical
investigation demands a nonperturbative approach
and, presently,
a pure QCD description of $\sigma_{tot}(s)$, from first principles and valid in the whole
energy region with available data, is still missing. 
In the phenomenological context $\sigma_{tot}(s)$ is usually investigated through
analytic parameterizations provided by the Regge-Gribov 
framework and fits to the experimental data available. 
In this formalism, the energy dependence of the total cross-section in the $s$-channel
is dictated by the kind of singularities of the amplitude in the $t$-channel,
typically poles (simple and triple). However, it should be noted that, in what concerns \textit{soft scattering states
of hadrons}, in particular, elastic scattering and therefore $\sigma_{tot}$, 
the Regge-Gribov approach is a kind of \textit{effective theory},
without an exclusive connection with QCD \cite{land}
(see also discussions on soft and hard Pomerons in \cite{pred},
Sect. 8.11).

In this situation, empirical analyses, able to describe the experimental data and,
simultaneously, looking
for connections with  formal and theoretical concepts, may constitute an important strategy
in the study of $\sigma_{tot}(s)$. In this direction, the 
log-raised-to-$\gamma$ component has shown to be a useful \textit{empirical tool} in the investigations.
Besides providing detailed \textit{local} information on the rise
of the total cross-section with the energy (and on the rates of changes, as slope and curvature),
it also allows useful checks on the FM upper bound, as well as, extrapolations to higher energies that 
may be important in cosmic-ray experiments.

In Appendix B, we have illustrated the efficiency and practical applicability of the RRPL2 and RRPL$\gamma$ models
in the analysis of $\sigma_{tot}$ data from $pp$ and $\bar{p}p$ above 5 GeV.
From the results and discussions, we may conclude that the data reductions here developed favor the RRPL$\gamma$
model, which indicates a rise of the total cross-section faster than
the log-squared dependence at the LHC energy region, namely $\gamma$ in the interval $\approx$ 2.1 - 2.4,
a result in accordance with
other analyses restricted to total cross-section
data \cite{velasco,fms1,fms2,ms1,ms2}. We have also quoted amplitude analyses, with simultaneous
fits to $\sigma_{tot}$ and $\rho$ data, predicting $\gamma$ values below 2 \cite{pdg14,pdg16,fms17b}.
The main point is to note that, 
since amplitude analyses with the exponent in the leading logarithmic component of $\sigma_{tot}(s)$  
treated as a free fit parameter have favored \textit{real} (\textit{not integer}) \textit{and positive} $\gamma$ \textit{values}, 
the associated singularity may not be a triple pole (which is associated with the leading
L2 law).

Based on the aforementioned arguments, in this work, as a first step in the search for possible connections between the empirical
L$\gamma$ law and formal concepts,
we have developed a \textit{mathematical study} on the singularities that could be
associated with this empirical leading component of $\sigma_{tot}$. 
The novel mathematical results consist of generalizations
of Eqs. (\ref{poles}) and (\ref{logn}), of integer order in the derivatives, to  
Eqs. (\ref{branch}) and (\ref{logr}), of arbitrary (real) order in the derivatives.
In this context, the dependence $\ln^{\gamma}(s/s_0)$ (with $\gamma > 0$ real, not integer) in the total cross-section is
mathematically associated
with a branch point, given by Eq. (\ref{bp}).

In what regards possible connections with the Regge-Gribov formalism,
we have noted that this branch point cannot be identified with 
Regge cuts, associated with multiple
Reggeons (or Pomeron) exchanges and treated through perturbative techniques.
On the other hand, our mathematical approach allow to infer some
novel interpretations that may be useful for further investigations,
as summarized in what follows.

Firstly, in Sect. 2, the use of the Mellin transform has established the connections between
asymptotic expansions and singularities \cite{bbo,joubert}, represented
by Eqs. (\ref{canonical}) and (\ref{singmellin}), respectively. In Sect. 3 the same analytic results
have been obtained through the fractional calculus, Eqs. (\ref{branch}) and (\ref{logr}), allowing, 
therefore, \textit{a fractional calculus
interpretation of the corresponding asymptotic expansions and singularities}
(demonstrated through the Mellin transform).
In this respect, we have argued that
the \textit{nonlocal character} of the Caputo fractional derivative,
replacing poles (integer order) by branch points (real order),
may bring associated additional information related
to memory, non-locality and interpolating effects
(between poles), which may point toward the need  of some reformulation.

Now, if we attempt to translate these fractional calculus interpretations to physical
arguments, the only candidates seem to be some missing nonperturbative effects,
which, obviously, is nothing new. However, we have shown that the \textit{real, positive and not integer}
$\gamma$-values, suggested by the phenomenological analyses
\cite{pdg14,pdg16,amaldi,ua42,velasco,fms1,fms2,ms1,ms2,fms17a,fms17b},
 demand a branch point singularity in the complex $J$-plane, Eq. (\ref{bp}), namely
an \textit{analytic result}, with fractional calculus interpretations, that may contribute with the search for useful
calculational schemes in nonperturbative QCD.

We expect that the results here presented, discussed and
interpreted (even if still based on some conjectures), may provide 
useful insights for future investigations, opening 
space for other aspects to be explored.

\section*{Acknowledgments}

Work supported by 
S\~ao Paulo Research Foundation (FAPESP),
Contract 2013/27060-3 (P.V.R.G.S.)


\appendix

\section{Total Cross Section and the Regge-Gribov Formalism}
\label{saa}

In this Appendix, after recalling the concept of total cross-section and referring
to the experimental data presently available at the highest energies, we proceed with a short
review on some basic concepts of the Regge-Gribov formalism.

\subsection{Total Hadronic Cross Section}

In high energy particle collisions, the total cross section is operationally defined by \cite{pred}
\begin{eqnarray}
\sigma_{tot} = \frac{N_{el} + N_{inel}}{\mathcal{L}}, 
\nonumber
\end{eqnarray}
where $N_{el}$ and $N_{inel}$ are the rate of elastic and inelastic interactions (scattered fluxes), respectively
and $\mathcal{L}$ is the luminosity (flux per unit area). 
This definition implies in both a statistical interpretation (probability of interaction
associated with the ratio between incident and scattered particles) and a geometrical
interpretation (related to the dimension, as an effective area of interaction).
For high-energy hadronic interactions $\sigma_{tot}$ is usually measured in millibarn (mb). 

For particle-particle and antiparticle-particle collisions, the largest interval
in energy with available data concerns proton-proton ($pp$) and antiproton-proton ($\bar{p}p$) scattering.
For further reference (Appendix B and main text), the experimental data presently available on $\sigma_{tot}$, from accelerator
experiments on $pp$ and $\bar{p}p$ collisions, as a function of the energy in the center-of-mass system,
$\sqrt{s}$, above 5 GeV, are displayed in Fig. 1. The highest energies with $pp$ collisions has
been reached at the CERN Large Hadron Collider (LHC): 7 TeV and 8 TeV.
The dataset, compiled from \cite{pdgdata}, includes the most recent measurements, at 8 TeV, by the TOTEM 
Collaboration \cite{totem} and 
ATLAS Collaboration \cite{atlas}.

\begin{figure}[ht]
\centering
\epsfig{file=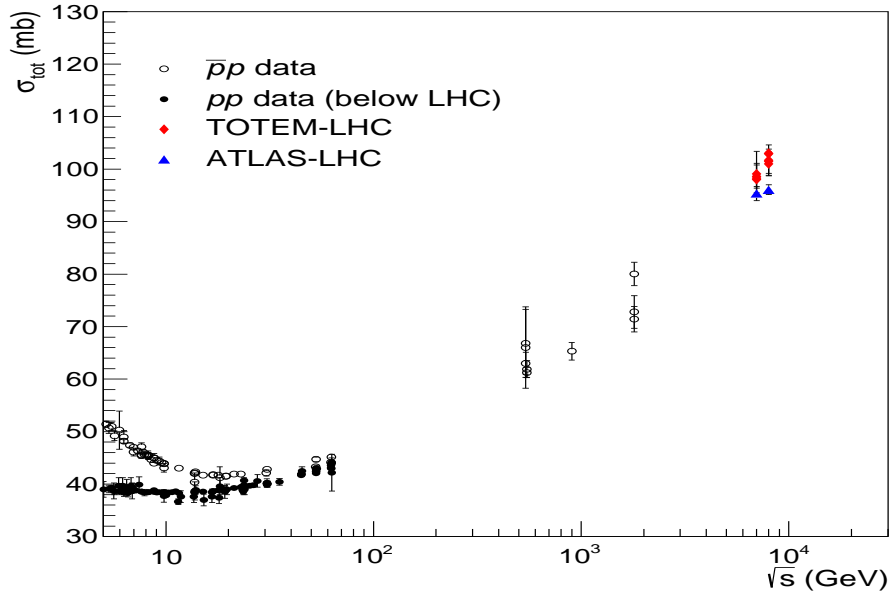,width=12cm,height=8cm}
\caption{Total cross-section data from accelerator experiments on $pp$ (black circles, diamonds and triangles) and 
$\bar{p}p$ (white circles) collisions at the LHC energies and below (compiled from \cite{pdgdata}).}
\label{f1}
\end{figure}

\subsection{Basic Phenomenological Concepts}

As commented in our introduction, amplitude analyses, based on the
$S$-Matrix formalism and Regge-Gribov theory \cite{pred,land,collins}, constitute the usual way to investigate
the behavior of the total hadronic cross-section. In this context, the total cross-section is
expressed as a sum of two terms, associated with Reggeons ($R$) and Pomerons
($P$) contributions. 
\begin{eqnarray}
\sigma_{tot}(s) = \sigma_{tot}^{R}(s) + \sigma_{tot}^{P}(s).
\label{rp}
\end{eqnarray}
Detailed treatment can be found in the
references quoted above and for a recent short review, see Appendix B in \cite{fms17b}.

The essential idea of the formalism is to associate the asymptotic behavior of the scattering
amplitude in terms of $s$ (named $s$-channel) with the singularities in the complex angular
momentum plane $J$ (named $t$-channel), $s$ and $t$ being the Mandelstam variables. In this context, 
a simple pole located at $J = \alpha(t)$ (a Regge Pole),
\begin{eqnarray}
\frac{1}{(J - \alpha(t))},
\nonumber
\end{eqnarray}
gives rise, for the asymptotic amplitude, to a power function of the energy \cite{pred,land,collins,fms17b},
\begin{eqnarray}
\mathrm{Im} \mathcal{A}(s, t) \propto s^{\alpha(t)}
\label{ima}
\end{eqnarray}
and therefore, at $t=0$, for $\sigma_{tot}(s)$ through Eq. (\ref{ot}).

The connection with particle exchanged is interpreted as Reggeons ($R$) exchanges, 
associated with the highest interpolated mesonic trajectories provided by
spectroscopic data ($t$-channel), relating Re$J = \alpha$ with
the masses $M^2$ (the Chew-Frautschi plot) \cite{pred}.
The trajectories are approximately linear, 
\begin{eqnarray}
\alpha(t) \approx \alpha_0 + \alpha' t,
\nonumber
\end{eqnarray}
defining an effective slope, $\alpha'$, and an intercept, $\alpha_0$,
which provides for $\sigma_{tot}$  ($t=0$) the exponent in the power law, negative and around 0.5
(Fig. 5.6 in \cite{pred}).
In the 1960s, with experimental data available only in the region
$\sqrt{s} <$ 20 GeV (see Fig. \ref{f1}), these dependencies could explain the decrease of 
$\sigma_{tot}(s)$ as the energy increases. Taking account of the crossing symmetry
and charge conjugation, these Reggeon contributions for $pp$ and $\bar{p}p$
scattering are expressed by
\begin{eqnarray}
\sigma_{tot}^R(s) &=& a_1 \left[\frac{s}{s_0}\right]^{-b_1} + \tau a_2 \left[\frac{s}{s_0}\right]^{-b_2}, 
\label{regge}
\end{eqnarray}
where $a_i, b_i, i = 1,2$ are real positive parameters, $s_0$ is a fixed energy scale and $\tau = -1$ for $pp$ and 
$\tau = +1$ for $\bar{p}p$. 

As new data have been obtained, new terms have been introduced. Firstly, the possibility
that $\sigma_{tot}(s)$ decreased to a constant value (not zero) led to the introduction
of a constant term, the Pomeranchuk constant or \textit{critical} Pomeron (see \cite{bor} for historical fits with
this contribution): 
\begin{eqnarray}
\sigma_{c}^P = A\ (\mathrm{constant}).
\end{eqnarray}
After the discovery of the rise of the total cross-section,
above $\approx$ 20 GeV (Fig. 1), an \textit{ad hoc} trajectory has been introduced with
positive intercept $\alpha_0$ slightly greater than 1, named \textit{supercritical} or \textit{soft} Pomeron
\cite{pred,land}, associated with a power law with exponent $\epsilon = \alpha_0 - 1$ slightly greater than 0:
\begin{eqnarray}
\sigma_{sc}^P = s^{\epsilon}.
\end{eqnarray}

However, since this dependence eventually violates the FM bound, other forms have been
studied, as logarithmic and/or logarithmic squared laws, which are associated with poles of
higher order, double and triple poles, respectively, as we show in Sect. 2.2 and 3.1.

In the general case, associated with a pole of order $N$ ($t$-channel), the contribution to the \textit{forward} amplitude
in the $s$-channel
is $s^{\alpha_0} \ln^{N-1}(s)$. In the case of the Pomeron, with a pole at $J = \alpha_0$ 
the contribution to the
total cross-section is 
 \begin{eqnarray}
\sigma^P_{tot}(s) = \frac{\mathrm{Im} \mathcal{A}(s,t=0)}{s} \propto s^{\alpha_0 - 1}\ln^{N-1} s.
\nonumber
\end{eqnarray}

Summarizing, we have the  associations:

\begin{itemize}

\item
simple pole ($N=1$) at $J = \alpha_0$, with $\alpha_0 = 1$\ \ $\Rightarrow$\ \ $\sigma^P_{tot}$ \ constant;

\item
simple pole ($N=1$) at $J=\alpha_0$\ \ $\Rightarrow$\ \ $\sigma^P_{tot} \propto s^{\alpha_0-1}$;

\item
double pole ($N=2$) at $J=\alpha_0$, with $\alpha_0=1$\ \ $\Rightarrow$\ \ $\sigma^P_{tot} \propto \ln(s)$;

\item
triple pole ($N=3$) at $J=\alpha_0$, with $\alpha_0=1$\ \ $\Rightarrow$\ \ $\sigma^P_{tot} \propto \ln^2(s)$.

\end{itemize}

\subsection{Analytic parameterizations}

\subsubsection{RRPL2 Model}

All the aforementioned different Pomeron contributions to $\sigma_{tot}(s)$ have been analyzed in detail by
the COMPETE Collaboration in 2002, by means of simultaneous fits to $\sigma_{tot}$
and $\rho$ data (ratio between the real and imaginary parts of the forward amplitude),
from several particle collisions ($pp$, $\bar{p}p$, mesons-$p$, $\gamma-p$, $\gamma-\gamma$).
Analytic connections between  $\sigma_{tot}$ and $\rho$ can be obtained in the context
of the Regge theory (power laws) or mathematical approaches, such as dispersion relations or
asymptotic uniqueness \cite{fms17b}. 
This study selected as the best model the one with
the Pomeron ($P$) contribution expressed by a constant critical Pomeron term plus a
log-squared contribution (triple pole) at the highest energies \cite{compete1,compete2}

\begin{eqnarray}
\sigma^{P}_{tot}(s) = A + B \ln^2\left(\frac{s}{s_0}\right),
\label{l2}
\end{eqnarray}
where $A$ and $B$ are free fit parameters. With the Reggeons contributions, Eq. (\ref{regge}),
the analytic result is expressed by the sum of the contributions
and is referred to as a RRPL2 model, standing for two Reggeons, Eq. (\ref{regge}), a constant
Pomeranchuk term and a triple pole Pomeron term (L2), Eq. (\ref{l2}).
This parametrization became a standard
result in forward amplitude analyses and has been used and updated
by the COMPAS group (IHEP, Protvino) in the
successive editions of the Review of Particle Physics, by the PDG \cite{pdg14,pdg16}.

\subsubsection{RRPL$\gamma$ Model}

In 1977 an \textit{empirical ansatz} was
introduced by Amaldi et al. \cite{amaldi} in the parametrization for $\sigma_{tot}$,
represented by a log-raised-to-$\gamma$ (L$\gamma$), with $\gamma$ a free fit parameter,
in place of the log-squared term (L2). From the above discussion, the full parametrization, recently denoted RRPL$\gamma$
in \cite{fms17a}, is expressed by
\begin{eqnarray}
\sigma_{tot}(s) = a_1 \left[\frac{s}{s_0}\right]^{-b_1} + \tau a_2 \left[\frac{s}{s_0}\right]^{-b_2} +
A + B \ln^{\gamma}\left(\frac{s}{s_0}\right), 
\label{lgamma}
\end{eqnarray}
where $a_i, b_i, i = 1,2$, $A$, $B$, $\gamma$ are free fit parameters, $s_0$ is a fixed energy scale and $\tau = -1$ for $pp$ and 
$\tau = +1$ for $\bar{p}p$. 

In this case ($\gamma$ as a real parameter), the connections between  $\sigma_{tot}$ and $\rho$ can be 
obtained either through numerical methods (integral dispersion relations) \cite{amaldi,ua42,velasco} or analytic
methods: derivative dispersion relations (DDR) \cite{fms1,fms2,ms1,ms2,fms17a,fms17b} or asymptotic uniqueness
(AU), which is based on the Phragm\'en-Lindel\"off theorems \cite{pdg14,pdg16,fms17b}. 
All the analyses employing dispersion relations favour central values of $\gamma$ above 2 (in the interval
2.1 - 2.6, depending on the data then available and the kind of fit: only $\sigma_{tot}$ data
or both $\sigma_{tot}$ and $\rho$ data). Contrasting with these results, analyses using the AU
method favor $\gamma$-values slightly below 2 \cite{pdg14,pdg16}
(see \cite{fms17b} for a detailed and comprehensive discussion on this subject).
Anyway, all these analyses lead to good and consistent descriptions of the experimental data
investigated. In the next Appendix, we illustrate the applicability of parametrization
(\ref{lgamma}) to a particular case.

\section{Data Reductions to $pp$ and $\bar{p}p$ Total Cross-Sections}
\label{sab}

In this Appendix, in order to stress the practical importance of the exponent $\gamma$
as a \textit{real} free parameter, we develop fits with the RRPL$\gamma$ model,
Eq. (\ref{lgamma}), to the total cross-section data presently
available at high energies (namely, above $\sqrt{s}$ = 5 GeV), shown in Fig. \ref{f1}.
For comparison, we shall also consider the particular case in which $\gamma = 2$ is fixed,
namely a RRPL2 model.

\subsection{Fit Procedures and Results}

As in our recent analysis
\cite{fms17a,fms17b}, 
we consider the energy scale fixed at the threshold for the
scattering states, namely $s_0 = 4m_p^2 = 3.521$ GeV$^2$, where $m_p$ is the
proton mass and
we initialize our parametric set
with the central values obtained in the most recent analyses by the PDG \cite{pdg16},
also displayed in \cite{fms17a}. 
With these initial values, we first develop the fit
with the RRPL2 model and using the final values as new feedbacks we develop the fit 
with the RRPL$\gamma$ model, namely by also letting free the parameter $\gamma$.

The data reductions were performed
with the objects of the class TMinuit of ROOT Framework 
\cite{root} and using the default MINUIT error analysis \cite{minuit}.
The error matrix provides the variances and covariances associated with each free parameter,
which are used in the analytic evaluation of the uncertainty regions 
associated with the fitted and predicted
quantities (through standard error propagation procedures \cite{bev}).
As tests of goodness-of-fit we shall consider the chi-square per degree of freedom
($\chi^2/\nu$) and
the corresponding integrated probability, $P(\chi^2)$ \cite{bev}.

The results of the fits are displayed in Table \ref{t1} (values of the free parameters
with the uncertainties and statistical information) and Fig. 2. 
Although not taking part in the data reductions, we included in this figure,
as illustration and for further reference, two estimations of the $pp$ total cross-section
from cosmic-ray experiments, beyond the accelerator energy region: 57 TeV (Auger) 
\cite{auger} and 95 TeV (Telescope Array) \cite{ta}.
The discrepancies between the TOTEM and ATLAS data (see Fig. 1) have
been recently discussed
in \cite{fms17a}.

\begin{table}[ht]
\centering
\caption{Fit results to accelerator data on $pp$ and $\bar{p}p$ scattering above $\sqrt{s}$ = 5 GeV with models 
RRPL2 and RRPL$\gamma$, Eq. (\ref{lgamma}).
Energy scale fixed, $s_0 = 4m_p^2 = 3.521$ GeV$^2$.} 
\vspace{0.1cm}
\begin{tabular}{c c c}
\hline
Model:        & RRPL2               & RRPL$\gamma$      \\\hline
$a_1$  (mb)   & 31.82 $\pm$ 0.70    & 31.2 $\pm$ 1.2    \\
$b_1$         & 0.398 $\pm$ 0.017   & 0.500 $\pm$ 0.077 \\
$a_2$  (mb)   & 16.72 $\pm$ 0.80    & 16.78 $\pm$ 0.84  \\
$b_2$         & 0.539 $\pm$ 0.015   & 0.540 $\pm$ 0.016 \\
$A$ (mb) & 29.93 $\pm$ 0.35    & 33.3 $\pm$ 1.8    \\
$B$  (mb) & 0.2455 $\pm$ 0.0028 & 0.130 $\pm$ 0.055 \\
$\gamma$      & 2 (fixed)           & \textbf{2.21 $\pm$ 0.14}   \\\hline 
$\nu$         & 168                 & 167               \\
$\chi^2/\nu$  & 1.02                & 1.01              \\
$P(\chi^2)$   & 0.420               & 0.457             \\
\hline 
\end{tabular}
\label{t1}
\end{table}

\begin{figure}[ht]
\begin{center}
 \includegraphics[scale=0.4]{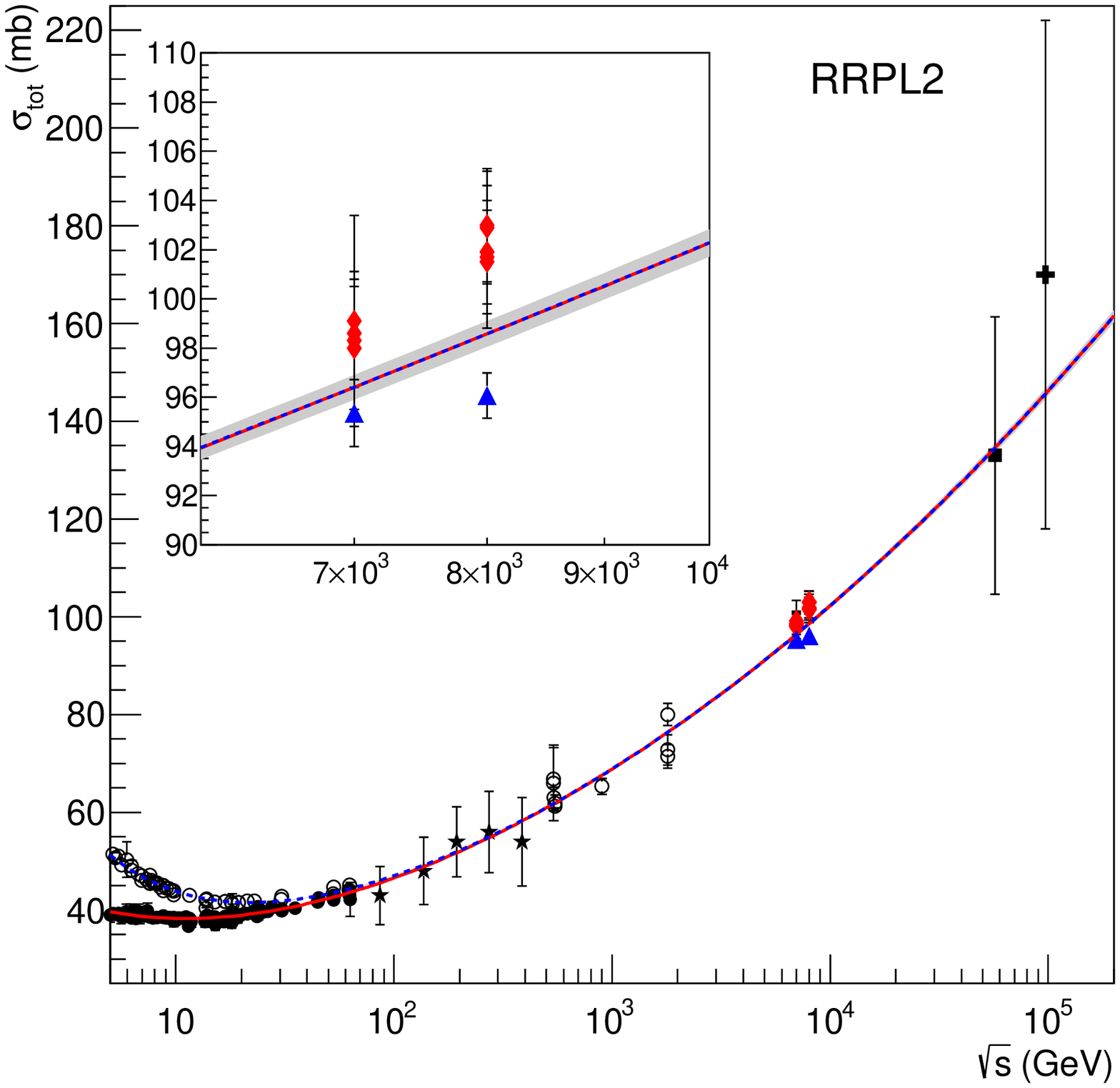}
 \includegraphics[scale=0.4]{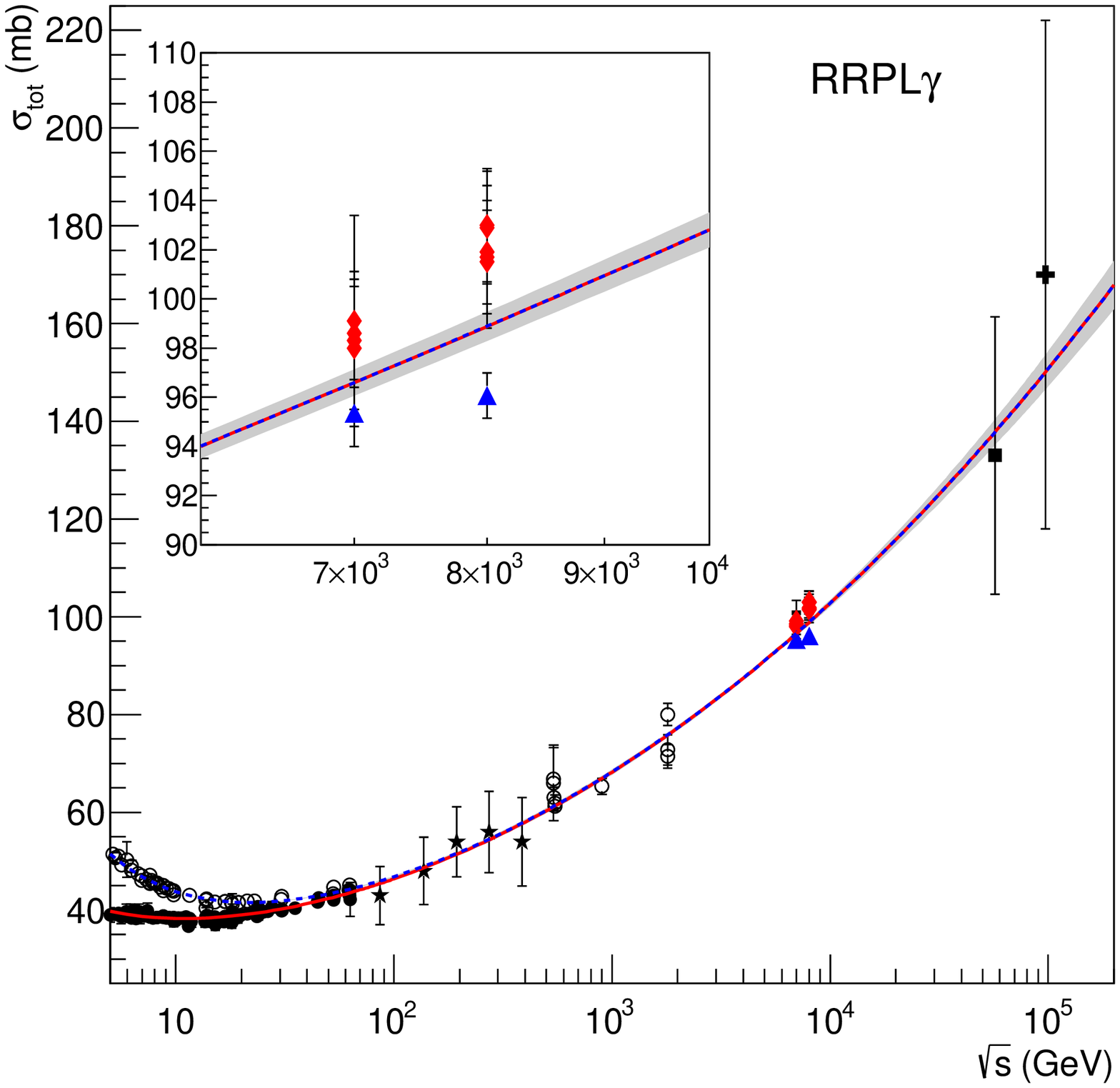}
 \caption{Fit results for the total cross-section, within the corresponding uncertainties regions,  
obtained through the RRPL2 and RRPL$\gamma$ models,
Eq. (\ref{lgamma}), Table \ref{t1}.
The insert shows the $\sigma_{tot}$ data and curves in the interval 6 - 10 TeV
(LHC region).}
\label{f2}
\end{center}
\end{figure} 

\subsection{Discussion on the Fit Results}

From Table 1, we see that the quality of the fit is good in both cases ($\chi^2/\nu \approx$ 1.0, for 
$\approx$ 170 degrees of freedom),
with a result slightly better, as expected, in case of $\gamma$ as a free fit parameter:
$P(\chi^2) \approx$ 0.4 (RRPL2) and
$P(\chi^2) \approx$ 0.5  (RRPL$\gamma$).

Although the curves are very similar,
note that differences appear as the energy increases above the accelerator data; indeed,
the prediction with the RRPL2 model reaches the central value of the cosmic-ray point
at 57 TeV (black square) and in case of the RRPL$\gamma$ model the prediction lies slightly above that central value,
within the uncertainties.
Once treated as a real free fit parameter, the value of $\gamma$ extracted from the fit,
2.21 $\pm$ 0.14, is consistent, within the uncertainties, with a rise of the total cross
section faster than the log-squared dependence.

It is important to note that
the \textit{leading high-energy component}, expressed by
\begin{eqnarray}
\sigma_{lead} = B \ln^{\gamma}(s/s_0),
\label{lead}
\end{eqnarray} 
provides also some interesting geometrical information. Indeed,
in terms of the variable $\ln(s/s_0)$, the rates of changes associated with
the above leading component, namely slope ($S$) and curvature ($C$), 
are given by
\begin{eqnarray}
S = B \gamma \ln^{\gamma -1}(s/s_0),
\qquad
C = B \gamma [\gamma - 1] \ln^{\gamma -2}(s/s_0).
\label{sc} 
\end{eqnarray} 
The RRPL2 model predicts constant curvature, $C_{L2} = 2 B$ and from Table 1,
we obtain $C_{L2} \approx$ 0.49 mb. On the other hand, in case of $\gamma > 2$,
the curvature increases as the energy increases. From Table 1,
$C_{L\gamma} \approx 0.3476 \ln^{0.21}(s/3.521)$ mb. For example, for $\sqrt{s} =$  10 TeV (typical of LHC region),
we obtain $C_{L\gamma} \approx $ 0.63 mb. 

We conclude that the data reduction with the RRPL$\gamma$ model
does indicate a rise of the total cross-section faster than the log-squared dependence,
as obtained in previous analyses restricted to total cross-section
data \cite{velasco,fms1,fms2,ms1,ms2} and suggested by some analyses including the $\rho$ data
through dispersion relations
\cite{amaldi,ua42,velasco,fms2,ms1,ms2,fms17a,fms17b}.
In other words, and for our purposes, these results \textit{disfavor
a triple pole singularity} (for which, $\gamma$ = 2).


\end{document}